\documentclass[smallextended,natbib]{svjour3}
\smartqed  
\usepackage{graphicx}
\usepackage{amsmath}
\usepackage{mathptmx}
\usepackage{latexsym}
\usepackage{bm}
\newcommand{\tr}{^{\prime}}

\def\bl#1{\mbox{\footnotesize \boldmath {$#1$}}} 
\def\cg#1{\mbox{${\cal #1}$}}      
\def\cgl#1{\mbox{\scriptsize {${\cal #1}$}}}
\begin{document}

\title{The Marginal Likelihood of two-way tables and Ecological Inference}
\titlerunning{Marginal Likelihood}        

\author{Antonio Forcina
}

\institute{A. Forcina \at Department of Economics, University of Perugia,
              via Pascoli 12, 06100, Italy \\
              \email{forcinarosara@gmail.com}           
}

\date{Received: date / Revised: date}

\maketitle

\begin{abstract}
The paper derives new results on the marginal likelihood of a two-way table which clarify the conditions under which Ecological inference is possible and lead to an efficient algorithm for maximizing the exact multinomial likelihood. The first part generalizes the work of \cite{Plackett77} on the marginal likelihood of a $2\times 2$ table to a general $R\times C$ table. In doing so, new conceptual tools are introduced and new insights on the geometry of the collection of tables having fixed row and column margins and the extended hypergeometric distribution are derived.
In the second part, when observations on the row and the column marginal distributions are available for a collection of two-way tables sharing the same association structure, an efficient Fisher scoring algorithm for maximizing the exact likelihood under multinomial sampling is introduced and a small simulation study is used to compare the performance of the proposed method with two well established ones.

\end{abstract}
The issue of whether it is possible to draw inference on the association structure of a two-way contingency table having observed only the row and column margins is relevant, for instance, in the context of ecological inference \citep[see, for instance, ][]{BrownPayne, Greiner2009, Beh2024}, a methodology which may be used to infer the distribution of voter decisions in a new election conditionally on the decisions taken by the same voters in an earlier election. Knowledge of such conditional distributions allow to estimate, for instance, the proportion of faithful voters of a given party or that of those who moved to a protest party or to abstention, see \cite{ForcinaPavia}.

Though, due to the secrecy of the ballots, the joint distribution of voters' decisions in two close in time elections cannot be observed, in many countries, the marginal distributions of voters recorded in each election, at each polling station, are available from official sources. Usually, voting options available in the earlier election are arranged by row
and those in the second election by column and the objective of inference are the conditional probabilities by row, that is the probability of voting option $j$ in the new election having voted option $i$ in the previous one.

Ecological inference is usually applied to data from a collection of local units, like polling stations, under the assumption that they share the same association structure. However, for instance, the information theoretic approach of \cite{Judge04} for general two-way tables, can be applied to estimate the conditional distributions from the row and column totals of a single table and the question is whether the resulting inferences are meaningful.
\cite{Plackett77} studied the log-likelihood of a single $2\times 2$ table when only the margins can be observed and noted that statements derived from likelihood based inference were "inconclusive".

In the first part of the paper we study the marginal likelihood for a single $R\times C$ contingency table and show that the likelihood has a collection of local maxima achieved, in the limit, as the log-odds ratios which are not 0 diverge towards the boundary of the parameter space. It turns out that the collection of probability distributions approaching a local maximum have an association structure such that, after suitable permutations of the row and the column categories, the log-odds ratios which are not equal to 0 tend to $+\infty$. In other words, there exists a permutation of the row and column categories such that the table achieving a specific local maximum is the table with the strongest possible positive association within the Fr\'echet class of bivariate distributions having the given pair of margins. In the concluding section of the paper we argue that this is an instance where maximum likelihood estimates are inconsistent.

Having established that it is essentially impossible to infer the conditional row distributions from the row and column totals of a single two-way table, in the second part of the paper we study the marginal likelihood for a collection of two-way tables assumed to share the same set of row conditional probabilities.
An efficient Fisher scoring algorithm which maximizes the exact multinomial marginal likelihood is introduced and a small simulation study is is used to show that the resulting estimates are close to the assumed true generating mechanism and perform better than two well established method of Ecological Inference.

Maximizing the exact multinomial likelihood has to face numerical complexities due to the need to construct, within each local unit, the collection of all possible tables compatible with the observed row and column totals; the size of these collections grow quickly with the number of observations and the number of row and column categories. However, once these collections have been constructed and stored, computation of the marginal likelihood and the conditional expectations according to the extended hypergeometric distribution are straightforward.

In section 2 we describe a parametrization of the conditional row distributions in the multinomial context, write the likelihood function and its derivatives, characterize the collection of distributions achieving local maxima, describe their properties and show that they satisfy the likelihood equations in the limit as certain log-odds ratios diverge. The extension to a collection of two-way tables sharing the same set of row conditional distributions is studied in section 3 where we derive the likelihood equations, determine a necessary condition for the information matrix to be non singular, propose a Fisher scoring algorithm for maximum likelihood estimation and present a small simulation study to compare the new method with two existing ones. Some general issues are discussed in section 4.
\section{The marginal likelihood of a single table}
\subsection{Notations and preliminary settings}\label{params}
Consider an $R\times C$ contingency table based on categorical variables, let $n_{ij}$ ($i=1,\dots ,R$, $j=1\dots ,C$) denote the joint frequencies, and assume that the marginal frequencies, $n_{i0},\: n_{0j}$, are strictly positive. Let $p_{ij}$ denote the joint probabilities, $p_{i0}$ = $\sum_j p_{ij}$ and $p_{j\mid i}$ = $p_{ij}/p_{i0}$ denote the conditional row probabilities. Conditional probabilities may be parameterized with logits and log-odds ration defined with reference to the initial category; more precisely, let $\phi_j$ = $\log[(p_{1j}/p_{11})]$, $\lambda_{ij}$ = $\log[p_{11}p_{ij}]$ - $\log[(p_{i1}p_{1j})]$ and note that $\phi_1=0$ and $\lambda_{ij}=0$ whenever $i=1$ and/or $j=1$; then
$$
p_{j\mid i} = \exp\left(\phi_j+\lambda_{ij}\right) / \sum_{j=1}^C\exp\left(\phi_j+\lambda_{ij}\right).
$$

Let $\bm N$ denote a table of joint frequencies and $\cg N$ the collection of all such tables compatible with the given row and column totals. The log likelihood may be computed by first summing across rows and columns and then across all possible configurations $\bm  N\in \cg N$. Let $\gamma(\bm N)$ = $-\sum_i\sum_j\log[\Gamma(n_{ij}+1)]$, $\bm\Lambda$ be the matrix with elements $\lambda_{ij}$ and $V(\bm N,\bm\Lambda)$ = $\sum_i\sum_j n_{ij} \lambda_{ij}$; after rearranging terms, we obtain
$$
L =\sum_{j=1}^C n_{0j} \phi_j + \log\sum_{\bm N\in\cg N} \exp\left(V(\bm N,\bm\Lambda) + \gamma(\bm N)\right) -\sum_{i=1}^R n_{i0}\log\sum_{j=1}^C\exp\left(\phi_j+\lambda_{ij}\right)
$$
%
\subsection{Solutions of the likelihood equations}\label{LE}
Let $M_{ij}$ denote the conditional expectation of $n_{ij}$ given the row and column totals,
it may be computed as
\begin{equation}
M_{ij} = \sum_{\bm N\in\cg N} n_{ij}\frac{ \exp\left(V(\bm N,\bm\Lambda) +
\gamma(\bm N)\right)}{\sum_{\bm N\in\cg N} \exp\left(V(\bm N,\bm\Lambda) + \gamma(\bm N)\right)}
\label{CondEx}
\end{equation}
By direct calculations, the likelihood equations may be written as
\begin{align}
n_{0j} &= \sum_{i=1}^R n_{i0}p_{j\mid i} \label{Leq1},\\
M_{ij} &= n_{i0} p_{j\mid i},\:\forall\:(i>1, \: j>1) \label{Leq2}.
\end{align}

Intuitively, as possible candidate solutions, one might consider the probability distributions $\cg P$ defined as follows: for each $\bm N\in \cg N$  use the unconstrained estimates  $\hat p_{j\mid i}=n_{ij}/n_{i0}$. It can be easily verified that these conditional probabilities satisfy likelihood equations (\ref{Leq1}). Inspection of likelihood equations (\ref{Leq2}) indicates that a $\bm P\in \cg P$ is a solution,  if and only if $M_{ij}=n_{ij}$ for all $i>1$, $j>1$.
In the next section we show that, though no element of $\cg N$ can satisfy (\ref{Leq2}), there exists a sub-collection, say $\cg Z \subset \cg N$, with the following property:  for each $\bm Z\in \cg Z$ we may construct a sequence of probability distributions $P(\bm Z,\epsilon)$, $\epsilon>0$, which satisfy (\ref{Leq2}) in the limit as $\epsilon\rightarrow 0$.
\subsection{Extreme tables}
Below we give a formal definition of the elements of $\cg Z$ by an algorithm for their construction, describe some of their properties and, for each extreme table, define a collection of row conditional probability distributions $P(\bm Z,\epsilon)$ and show that, as $\epsilon\rightarrow 0$, they satisfy likelihood equations (\ref{Leq2}). In short, an extreme table is the upper bound of the Fréchet class of distributions after applying a pair of permutations, say $\bm\pi_r,\:\bm\pi_c$ to the row and column categories, see \cite{Frechet1960}, equation (8) on p. 11 and \cite{CifaRega2017} for a short history of the subject and some properties of the joint distributions derived from extreme tables. An informal algorithm for their construction is described by \cite{Salvemini19439}, a disciple of C. Gini who apparently was aware of these tables before the work of  Fréchet.
\subsubsection{Definition and construction}
Let $\bm m_r$, $\bm m_c$ denote, respectively, the vectors with elements $n_{i0}$ and $n_{0j}$;
\begin{definition}
\label{MaxDef}
Given a pair of marginal distributions $\bm m_r,\:\bm m_c$, we call {\bf extreme} a frequency table with entries $z_{ij}$ obtained by the following procedure: (i) select a pair of permutations one for the row categories and one for the column categories, (ii) apply {\bf Algorithm 1} below.
\end{definition}

{\bf Algorithm 1}
{\it Construction of an extreme frequency table}\label{alg1}.
Let $\bm\pi_r,\bm\pi_c$ be two arbitrary permutations of the row and the column categories respectively, let the vector $\bm t_r=\bm m_r(\bm \pi_r)$ with elements $t_{i0}$ and $\bm t_c=\bm m_c(\bm \pi_c)$ with element $t_{0j}$: the algorithm may start either from cell (1,1) or $(R,C)$; the version starting from (1,1) is described below. The algorithm is a sequence of moves where each step determines the next cell to be processed:
\begin{itemize}
  \item  Start by setting $i=1,\:j=1$;
  \item  Once in cell (i,j):
  \begin{enumerate}
  \item  If $t_{i0}-\sum_{h=1}^{j-1} t_{ih} < t_{0j}-\sum_{h=1}^{i-1} t_{hj}$, set $z_{ij} = t_{i0}-\sum_{h=1}^{j-1} t_{ih} $, $z_{ik}=0$ $\forall k>j$ and go to cell (i+1,j);
  \item If $t_{i0}-\sum_{h=1}^{j-1}t_{ih} > t_{0j}-\sum_{h=1}^{i-1}t_{hj}$, set $z_{ij} = t_{0j}-\sum_{h=1}^{i-1}t_{hj}$, $z_{hj}= 0$ $\forall h>i$, go to cell (i,j+1);
  \item if $t_{i0}-\sum_{h=1}^{j-1} t_{ih} = t_{0j}-\sum_{h=1}^{i-1} t_{hj}$, set $z_{ij}=t_{i0}-\sum_{h=1}^{j-1} t_{ih}$, $z_{ik}=0$ $\forall k>j$ and $z_{hj}=0$ $\forall h>i$,
    move to cell $i+1,j+1$;
  \end{enumerate}
  \item Go back to (1) and continue until cell $(R,C)$ has been processed.
\end{itemize}
\begin{lemma} \label{Frechet}
The table $\bm Z$ produced by Algorithm 1 satisfies the right-hand side constraints in \cite{Frechet1960}, equation (8).
\end{lemma}
{\sc Proof}
When $t_{i0}-\sum_{h=1}^{j-1} t_{ih}$ and $t_{0j}-\sum_{h=1}^{i-1} t_{hj}$ are both greater than 0, $z_{ij}$ is set equal to the the smallest of these two expressions; $z_{ij}$ is set to 0 otherwise-
\begin{example}
\label{Ex1}
Suppose that the vectors of row and column totals are equal, respectively, to (44, 37, 57, 62) and
(57, 58, 42, 43), the three instances of extreme tables displayed in Table \ref{Tab1} are obtained by the following pairs of permutations:
(3 4 1 2), (2 1 3 4); (1 3 2 4), (2 4 1 3); (3 2 1 4), (2 3 4 1).
\begin{table}[h!]
\centering
	\caption{Three instances of extreme frequency distributions}.	
    \begin{tabular}{rrrrrcrrrrrcrrrrr}
	\multicolumn{5}{c}{$\bm Z_1$} &\hspace{1mm} & \multicolumn{5}{c}{$\bm Z_2$} &\hspace{1mm} &
\multicolumn{5}{c}{$\bm Z_3$} \\
    57 &  0 &  0 &  0 & 57 && 42 &  2 &  0 &  0 & 44 && 43 & 14 &  0 &  0 & 57\\
     1 & 57 &  4 &  0 & 62 &&  0 & 55 &  2 &  0 & 57 &&  0 & 37 &  0 &  0 & 37\\
     0 &  0 & 38 &  6 & 44 &&  0 &  0 & 37 &  0 & 37 &&  0 &  6 & 38 &  0 & 44\\
     0 &  0 &  0 & 37 & 37 &&  0 &  0 &  4 & 58 & 62 &&  0 &  0 & 20 & 42 & 62\\
    58 & 57 & 42 & 43 &200 && 42 & 57 & 43 & 58 &200 && 43 & 57 & 58 & 42 &200\\
	\end{tabular}
\label{Tab1}
\end{table}
\end{example}
\begin{remark}
If the row and column categories were ordinal and such that, after the rearrangement determined by $\bm\pi_r,\:\bm\pi_c$ they were both increasing, the resulting extreme table would be the table with the strongest positive association within all possible tables having the same marginal distributions. The table with the strongest negative association would be the extreme table where the order of the column categories has been reversed. In principle, the number of extreme tables is equal to the product of the possible permutations of the row and the column categories; however, the  actual number of distinct tables might be smaller because different pairs of permutations may lead to the same extreme table.
\end{remark}
\subsection{Main results} \label{MaxT}
Recall that each extreme table is determined by a pair of permutations of the row and the column categories, by construction, they contain several 0 entries where the corresponding log-odds ratios $\lambda_{ij}$ = $\log[z_{11}z_{ij}/(z_{i1}z_{1j})]$, are undefined. In the following, for conciseness, let $\bm Z$ = $\bm Z(\bm\pi_r,\bm\pi_c)$.

For any given $\bm Z$, we may define a sequence of tables $\bm Z(\epsilon)$, $\epsilon>0$, which converge to $\bm Z$ as $\epsilon\rightarrow 0$ and are such that the log-odds ratios are all well defined.  Table \ref{Tab2} displays the possible configurations of 0s in $2\times 2$ sub-tables which include the cell (1,1) and are thus relevant in the computation of the log-odds ratios.
\begin{table}[h!]
\centering
	\caption{Possible configurations of 0s in $2\times 2$ sub-tables of an extreme table that include the (1,1) cell}.	
\begin{tabular}{rrcrrcrrcrrcrrcrr}  \\ \hline
\multicolumn{2}{c}{$A_1$} &\hspace{0.8mm} & \multicolumn{2}{c}{$A_2$}  & \multicolumn{2}{c}{$A_3$}
&\hspace{0.8mm} &\multicolumn{2}{c}{$B_1$} &\hspace{0.8mm} & \multicolumn{2}{c}{$B_2$} &\hspace{0.8mm} &
\multicolumn{2}{c}{$C$} \\ \hline
$z_{11}$ & 0 && $z_{11}$ & $z_{1j}$ && $z_{11}$ & 0 && $z_{11}$ & 0         && $z_{11}$ & $z_{1j}$
&& $z_{11}$ & 0 \\
$z_{i1}$ & 0 && 0        & 0        && 0        & 0 && $z_{i1}$ & $z_{ij}$  && 0        & $z_{ij}$
&& 0 & $z_{11}$ \\ \hline
	\end{tabular}
\label{Tab2}
\end{table}

A procedure for approximating any given $\bm Z$ with a table $\bm Z(\epsilon)$ which converges to $\bm Z$ as $\epsilon\rightarrow 0$ and is such that the log-odds ratio are all well defined is obtained by replacing 0s in the first row and first column with $\epsilon$ and, for $Z_{ij}=0$ ($i>1,\: j>1$) do the following
\begin{itemize}
\item if $Z_{1j}(\epsilon)=0$ and $Z_{i1}>0$, as in $A_1$, set $Z_{ij}=\epsilon Z_{i1}/Z_{11}$;
\item if $Z_{i1}(\epsilon)=0$ and $Z_{1j}>0$, as in $A_2$, set $Z_{ij}=\epsilon Z_{1j}/Z_{11}$;
\item if $Z_{1j}(\epsilon)=0$ and $Z_{i1}=0$, as in $A_3$, set $Z_{ij}=\epsilon^2 /Z_{11}$;
\end{itemize}
Let $\lambda_{ij}(\epsilon)$, $i,j\: >1$, denote the log-odds ratios computed in $\bm Z(\epsilon)$ and $\xi=-\log(\epsilon)$. It can be easily seen that the log-odds ratios within any $\bm Z(\epsilon)$ will be 0 under A, of order $O(\xi)$ under B and of order $2O(\xi)$ under C. When $\epsilon$ is sufficiently close to 0, the $\lambda_{ij}(\epsilon)$ which are different from 0, tend to $\xi$ or $2\xi$. Construct $P(\bm Z,\epsilon)$ by dividing the elements of $\bm Z(\epsilon)$ by the corresponding row totals; clearly this does not affect the log-odds ratios.

The following Lemma states some basic properties of extreme tables and the corresponding $P(\bm Z,\epsilon)$.
\begin{lemma} \label{Lem1}
Suppose that the extreme table $\bm Z$ is determined by the pair of permutations $\bm\pi_r,\bm\pi_c$; let $\cg N^*$ denote the collection $\cg N$ after applying $\bm\pi_r,\bm\pi_c$ to each of its elements, then
\begin{description}
\item{(i)} If $\lambda_{ij}(\epsilon)$ is of order $O(2\xi)$, $\lambda_{hk}(\epsilon)$, $h\geq i$ and $k\geq j$,  is either 0 or of order $O(2\xi)$;
\item{(ii)} If $\lambda_{ij}(\epsilon)$ is of order $O(\xi)$,  $\lambda_{hk}(\epsilon)$, $h\geq i$ and $k\geq j$, is either 0 or, at least, of order $O(\xi)$.
\item{(iii)} for $\:i>1,\:j>1$ and any $\bm N^*\in \cg N^*$,
$$
\sum_{h=i}^R\sum_{k=j}^C z_{hk} \geq \sum_{h=i}^R\sum_{k=j}^C n_{hk}^*.
$$
\end{description}
\end{lemma}
{\sc Proof}
The construction algorithm of extreme tables implies that, if $z_{i1}=z_{1j}=0$, $\forall \:h>i,\:k>j$, $z_{h1}=z_{1k}=0$; it follows that sub-tables defined by the $(1,1$) and $(h,k)$ cells are either of type $A_3$ or $C$ implying (i). Concerning (ii), the corresponding sub-tables may be of type $B_1$ or $B_2$, in the first case $z_{1k}$, $k>j$ must be 0 while $z_{h1},\: z_{hk}$ may either be 0 or positive. The last point follows by constructing the extreme table after collapsing row categories 1 to $i-1$ and $i$ to $R$ and column categories 1 to $j-1$ and $j$ to $C$; the content of cell (2,2) in the resulting extreme table is the maximum allowed by the margins. $\Box$
\begin{lemma} \label{Lem2}
Let $\bm Z$ be an extreme table and $\bm\Lambda(\epsilon)$ be the matrix of log-odds ratios computed in $P(\bm Z,\epsilon)$ for any table of counts $\bm N^*\in\cg N^*$ and indices $s,t,u,v$ such that $2\leq s<u$ and $2\leq t<v$,
$$
V(\bm N^*,\bm\Lambda(\epsilon))-V(\bm Z,\bm\Lambda(\epsilon)) = \xi\left[\sum_{i=s}^R\sum_{j=t}^C
(n^*_{ij}-z_{ij}) + \sum_{i=u}^R\sum_{j=v}^C (n^*_{ij}-z_{ij}) \right]+o(\epsilon),
$$
in case no block of type C is present, the $u,\: v$ term may be dropped.
\end{lemma}
{\sc Proof}
Recall that $V(\bm N^*,\bm\Lambda)$ = $\sum_i\sum_j n^*_{ij} \lambda_{ij}$ and note that one can ignore the components of $V(\bm N^*,\bm\Lambda(\epsilon))$ with $\lambda_{ij}(\epsilon))=0$; to a first order of approximation, the remaining terms are either of order $O(\xi)$ or $2O(\xi)$, with the latter ones, if present, located towards the lower right-hand corner and the result follows by rearranging terms.$\Box$

The main result of this section is contained in the next Theorem:
\begin{theorem}
For any extreme table $\bm Z\in\cg Z$, the probability distribution $P(\bm Z,\epsilon)$ satisfies likelihood equations (\ref{Leq1}) and (\ref{Leq2}) in the limit as $\epsilon \rightarrow 0$ and is, thus, a local maximum of the marginal likelihood.
\end{theorem}
{\sc Proof}
Though, as $\epsilon$ tends to 0 the elements of $V(\bm N^*,\bm\Lambda)$, $\bm N^*\in (\cg N^*\backslash \cg Z)$ diverge as $\epsilon\rightarrow 0$, the conditional expectations in (\ref{CondEx}) may still be computed by subtracting the maximum of $V(\bm N^*,\bm\Lambda)$ across the collection of possible tables both from the numerator and the denominator. Then, (iii) from Lemma \ref{Lem1} applied to the statement in Lemma \ref{Lem2} implies that $V(\bm N^*,\bm\Lambda(\epsilon))-V(\bm Z,\bm\Lambda(\epsilon))$ converges to $-\infty$ as $\epsilon\rightarrow 0$ so that the corresponding exponentials tends to 0 and may be ignored, except for the exponential where $\bm N^*$ = $\bm Z$ which tends to 1; this implies that $M_{ij}$ = $n_{ij}$ for all $i,j$. $\Box$
\subsection{Final comments}
It can be easily seen that the likelihood equations are also satisfied by the table of conditional probabilities computed under independence which are determined by the  column totals, that is $p_{j\mid i}$ = $n_{0j}/n$:
$$
\sum_i n_{i0}n_{0j}/n = n_{0j}, \quad M_{ij} = n_{i0 }n_{0j}/n.
$$
In the $2\times 2$ case, \cite{Plackett77} showed that the likelihood function has a saddle-point at independence; though an extension of his argument to the $R\times C$ context is not straightforward, examination of several numerical examples determined by a pair of row and column totals such that the corresponding collection $\cg N$ is not too large to construct, indicates that the likelihood under independence is not larger than that of any $\bm N\in\cg N$; however there is a variety of probability distributions not derived from members of $\cg N$ whose likelihood is even smaller than under independence. An instance is presented in Example \ref{Ex2}.
\begin{example} \label{Ex2}
For each member of the collection of 2160 frequency tables compatible with the row margin (8,20,12) and  column margin (12,7, 21) a probability distribution was constructed by replacing 0s with $\epsilon=10^{-50}$ and scaling row totals to sum to 1. The log-likelihood of these probability distributions, arranged in the lexicographic order in which feasible tables are generated, are plotted in Figure \ref{F1}. The likelihood of extreme tables is marked with a square and the dotted line is the likelihood under independence. In addition, a collection of distributions obtained by mixing the independence distribution with conditional row distributions generated at random are also displayed; however, to improve visibility, among the distributions generated at random, only those whose likelihood was above a certain threshold are retained.

The likelihood under independence seems to be below that of any distribution derived from $\bm N\in \cg N$ while most of the mixture distribution have a likelihood below that under independence. Certain probability distributions derived from tables which are not extreme, may have a likelihood larger than that of several extreme tables. This happens when a non extreme table may be obtained by an elementary swap, as defined by \cite{Diaconis1998}, from  an extreme table with a very high likelihood. The kind of regular pattern appearing in the plot is due to the fact the algorithm that generates the collection $\cg N$  moves by elementary swaps so that successive tables are in some sese adjacent.
\begin{figure}[!h]
\begin{center}
\includegraphics[width=1.1\textwidth]{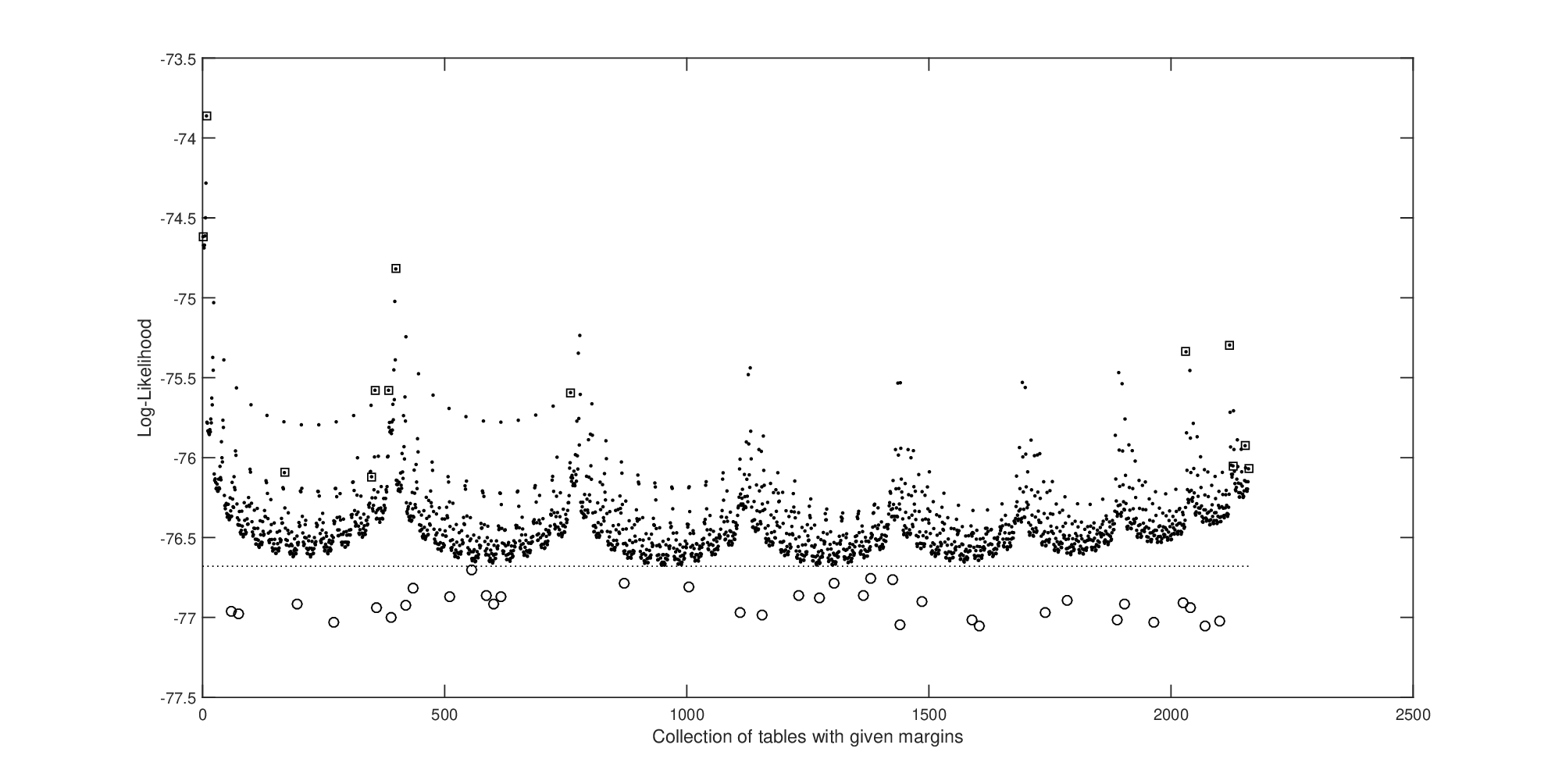}
\end{center}
\caption{Plots of the Log-likelihood, vertical axis, for all the distributions which may be derived from tables compatible with the row totals (8,20,12) and column totals (12,7,21); extreme distributions are marked with a square, ordinary ones with "$\cdot$"; the horizontal dotted line corresponds to independence and "$o$" to distributions obtained by mixing independence with conditional distributions generated at random.}
\label{F1}
\end{figure}
\end{example}

The probability distribution which achieves the global maximum is, obviously, derived from an extreme table; however, which extreme table leads to the global maximum is hard to say in general because, in addition to its log-odds ratios, the global maximum depends also on numerical features of the full collection  $\cg N$,
\section{Marginal likelihood for a collections of two-way tables}
\subsection{Introduction}
Suppose that row and column distributions have been observed for a collection of local units, each characterized by a two-way table, $\bm N_h$, $h=1,\dots , s$ whose content, for some reason, cannot be observed, except for the row and column totals, say $n_{hi0},\: n_{h0j}$, $i=1,\dots, R$, $j=1,\dots ,C$. In studies of voting behaviour in two close in time elections, $\bm N_h$ would be the cross tabulation of voters choices in the two elections in polling station $h$; clearly, this table cannot be observed due to the secrecy of the ballot. A similar issue arise in various contexts of Epidemiology, for a discussion of its consequences, see for instance \cite{Greenland1989}.

Methods of ecological inference under the assumption that the given collection of two-way tables share the same set of conditional row distributions have been used since the first half of last century, like in \cite{Gosnell1935}. Nonetheless it has always been looked at with suspicion within the sociological community, especially after \cite{Robinson} showed  that its estimates could be heavily biased. A formal statement of the problem was provided by \cite{Goodman} who stated the conditions under which the approach based on linear regression (which was current practice at the time) provides unbiased estimates of the conditional row probabilities which are the quantities of interest.

Methods of ecological inference are usually based on the assumption that the conditional probabilities $p_{hij}/p_{hi0}$ = $p_{j\mid i}$, that is they are constant across local units. Assume that these conditional probabilities are parameterised as in section \ref{params}; to write the log likelihood, let $n_{hi0}$ and $n_{h0j}$, $h=1,\dots,s$ denote respectively the row and column totals. Then log likelihood may be written as
$$
L =\sum_{j=1}^C n_{00j} \phi_j + \sum_{h=1}^s\log\sum_{\cg N_h} \exp\left(V(\bm N_h,\bm\Lambda) + G(\bm N_h)\right) -\sum_{i=1}^R n_{0i0}\log\sum_{j=1}^C\exp\left(\phi_j+\lambda_{ij}\right).
$$
By a straightforward extension of the argument in \ref{LE}, the likelihood equations have the simple form
$$
n_{00j} = \sum_{i=1}^R n_{0i0}p_{j\mid i},\quad  \sum_{h=1}^s M_{hij} = n_{0i0} p_{j\mid i},\:\forall i>1, \: j>1.
$$
While $\hat p_{j\mid i}$ = $\sum_h n_{hij}/n_{0i0}$ satisfies the first set of likelihood equations,
finding estimates that, in addition, satisfy the second set of likelihood equations is much harder because, for any candidate estimate $\hat p_{j\mid i}$, the table of conditional expectations from the extended hypergeometric distribution have to be computed numerically within each collection $\cg N_h$.

It is worth noting that, in order for the kind of pathological phenomenon which arise when a single pair of marginals is available, the $s$ pairs of marginals should have at least one common maximal table, which is impossible unless both row and column totals are proportional across polling stations.
\subsection{A new method of Ecological Inference}
A simple estimate of $\hat p_{j\mid i}$ may be computed by the linear regression method of \cite{Goodman} which is worth reviewing briefly. Let $\bm U,\:\bm V$ denote, respectively, the $s\times R$ and $s\times C$ matrices with elements $n_{hi0}/n_{h00}$ and $n_{h0j}/n_{h00}$, with $h$ indexing rows. Let $\bm X$ = $\bm U\otimes \bm I_{C-1}$ and $\bm y$ be the column vector obtained by arranging the matrix made of the first $C-1$ columns of $\bm V$ row-wise. Let $\hat{\bm \beta}$ = $(\bm X\tr \bm X)^{-1}\bm y$; an estimate of the $p_{j\mid i}$s is obtained by arranging the elements of $\hat{\bm \beta}$ into an $R\times (C-1)$ matrix $\bm A$ where $\hat{p}_{j\mid i}$ = $A_{ij}$ for $j<C$. Clearly,  $rank(\bm X)$ = $s$ if and only if $s\geq R$ and the rows of $\bm U$ are linearly independent. Thus, at least $R$ polling stations sharing the same row conditional distributions are needed for the method to be applied.

Limitations of the estimates by the Goodman's method have been discussed for instance by \cite{FoGnBr} and are summarized below:
\begin{itemize}
\item Because estimates from the regression model range on the whole real line, estimated probabilities may be negative or greater than 1;
\item The least square estimates use only the condition that $E(\bm y)$ = $\bm X\bm \beta$, but ignores variances and thus is not efficient.
\end{itemize}
We now outline a Fisher scoring algorithm for maximizing the exact multinomial likelihood;
it may be easily formulated and  implemented in practice as long as the collections $\cg N_h$, $h=1,\dots, s$, can be constructed in finite time and stored.

In order to use a matrix formulation, let $\bm\beta$ be the $R\times (C-1)$ vector obtained by stacking the elements of $\phi_j$ and then those of $\lambda_{ij}$ with $j$ running faster. Next define $\bm u_h$ to be the score vectors within each local unit; they have elements
$$
n_{h0j} - \sum_{i=1}^R n_{hi0}p_{j\mid i}
$$
in the first $C-1$ positions and
$$
M_{hij} - n_{hi0} p_{j\mid i},
$$
with $j$ running faster, in the remaining positions.

An alternative to the expected information matrix which, in this context, seems hard to compute, is the {\it empirical information matrix} $\bm E$, \citep{Scott2002} which is defined next. Let $\bm u$ = $\sum \bm u_h$ and $\bm S$ be the $R(C-1)\times s$ matrix collecting score vector for each single table, thus its $h$th column is equal to $\bm u_h$, next define
$$
\bm E = (\bm S - \bm u\bm u\tr/s) (\bm S - \bm u\bm u\tr/s)/(s-1).
$$

It can be shown, \citep{Forcina2017EcS}, that the the empirical information matrix is an unbiased estimate of the expected information matrix. In addition, $rank(\bm E)$ = $rank(\bm S)$, thus, a necessary condition for identifiability is that $s\geq R(C-1)$; this condition is also sufficient, except for degenerate data, like with the Goodman method.
\subsubsection{A small simulation study}

For $s=60$ local units, $R=C=3$ and the number of voters in each local unit set to $n=40$, artificial data were generated as follows. For each local unit, the row margins $n_{hi0}$ were sampled from a multinomial with uniform  probabilities  and saved into the rows of the $\bm U$ matrix; next, the results at election 2 within each local unit and each row were sampled from a multinomial of size $n_{hi0}$ and probabilities equal to the rows of the matrix $\bm\Pi$ given in Table 3. This procedure produces the matrices of joint frequencies $\bm N_h$, $h=1,\dots, s$ whose column totals $n_{h0j}$ are then saved into the $h$th row of $\bm V$.

With these data, in addition to the Fisher scoring algorithm described above, we computed estimates from the Goodman linear regression and by the modified Brown and Payne method as described in \cite{FoGnBr}. Linear regression estimates were forced within  the (0,1) range when outside and results adjusted by iterative proportional fitting and scaled back into conditional probabilities.

\begin{table}[!h]
\centering
	\caption{Tables of conditional transition probabilities: $\bm\Pi$ is the true model, $\hat{\bm\Pi}$ is the table estimated by the Fisher scoring algorithm based on  the 60 local units and $\bm\Pi_{\bl Z}$ is derived from the extreme table achieving the global maximum having row and column totals equal to the sum of the corresponding totals across local units.}	
 \begin{tabular}{rrrcrrrcrrr} \hline
	\multicolumn{3}{c}{$\bm\Pi$} &\hspace{3mm} & \multicolumn{3}{c}{$\hat{\bm\Pi}$} & \hspace{3mm} & \multicolumn{3}{c}{$\bm\Pi_{\bl Z}$} \\ \hline
    0.490 & 0.280 &  0.230 &&  0.435 & 0.313 & 0.252 && 0.941 & 0.059 & 0.000\\
    0.300 & 0.480 &  0.220 &&  0.346 & 0.450 & 0.204 && 1.000 & 0.100 & 0.000\\
    0.210 & 0.320 &  0.470 &&  0.245 & 0.312 & 0.443 && 0.000 & 0.000 & 1.000 \\ \hline
	\end{tabular}
\label{Tab3}
\end{table}

The accuracy of the different estimates was assessed by the following quantity
$$
M_e = \sqrt{\frac{1}{sRC}\sum_{h=1}^s\sum_1^R\sum_1^C (\hat P^{(e)}_{hij} -\Pi_{ij})^2}
$$
where $\Pi_{ij}$ are the true conditional probabilities and $\hat P^{(e)}_{ij}$ are the corresponding estimates obtained by method "e", that is exact likelihood, Brown and Payne and Goodman.
$M_e$ is equal to 0.0537 for maximum likelihood, to 0.0549 for the modified Brown and Payne and to  0.0551 for the Goodman method. The result indicates that, in spite of its limitations, the Goodman method does not perform much worse than the modified Brown and Payne method which is among the most sophisticated methods of ecological inference and assumes a mixture of Dirichelet multinomial distributions allowing for voters to influence each other within small circles. Because that model includes the plain multinomial as a special case it should perform at least as well. The result may be due to the fact that the modified Brown and Payne method uses a normal approximation to the true likelihood which may not be accurate due to the sample sizes used in the simulation.

Table \ref{Tab3} displays the true conditional probabilities, those estimated by maximizing the exact likelihood for the 60 local units and that corresponding to the global maximum if only the overall marginals for the 60 local units was available. While the estimates obtained by using the full data are rather close to the truth, the anomaly, so to speak, is that the estimates obtained by using the overall row and column totals are correspond to a highest positive association compatible with the given margins
\section{Discussion}
Though the main objective of Plackett when studying the marginal likelihood of a $2\times 2$ table was to clarify whether the marginal distributions contained some information about the degree of association between the row and column variables, in the abstract he notes that the indication provided by "standard procedures", that is likelihood inference, "are inconclusive" and in the final discussion seems to suggest that this might be one of the problems where the likelihood approach fails.

Below we provide an informal proof that maximum likelihood estimates based on observing the margins of a single $R\times C$ table are inconsistent. For simplicity suppose we have a $3\times 3$ table where the row margins have a uniform distribution and that the true row conditional probabilities are those in the matrix $\bm\Pi$ in Table \ref{Tab3}. The true column margins can be easily derived and are very close to uniform. Suppose that we search for the maximum likelihood estimates in samples of increasing size derived from such a model; though the resulting class of tables $\cg N$ is practically impossible to construct for samples of size much larger than 200, Theorem 1 implies that the global maximum of the likelihood will be attained in one of the 36 possible extreme tables; in the $3\times 3$ case these tables must have at least four cells with conditional probability equal to 0. For $n=100,000$ three instances of extreme tables are displayed in Table 3; though it is impossible to determine which table achieves the global maximum, when marginal distribution are close to uniform, the log-likelihood achieved by any $\bm Z(\epsilon)$ with $\bm Z\in \cg Z$ and $\epsilon\rightarrow 0$ are not much different. Suppose that we measure the discrepancy between estimates derived from extreme tables and true values a kind of mean square error, that is
$$
M=\sqrt{\frac{1}{9}\sum_{i=1}^{R}\sum_{j=1}^C (z_{ij}/z_{i0}-\Pi_{ij})^2},
$$
where $\Pi_{ij}$ are the elements of $\bm \Pi$ in Table 3 and the $z_{ij}$ are the elements of an extreme table;clearly $M$ is minimized by $\bm Z_1$ in Table \ref{Tab4} where $M=0.368$ and remains approximately constant when the sample size increases
\begin{table}[!h]
\centering
	\caption{Three instances of extreme tables whose likelihood is close to the global maximum for samples of size 100,000.}	
 \begin{tabular}{rrrcrrrcrrr} \hline
\multicolumn{3}{c}{$\bm Z_1$} &\hspace{3mm} & \multicolumn{3}{c}{$\hat{\bm Z_2}$} & \hspace{3mm} & \multicolumn{3}{c}{$\bm Z_3$} \\ \hline
  33193 &      0 &      0 &&      0 &      0 &  33193 &&  33193 &      0  &     0\\
    125 &  33233 &     48 &&    173 &  33233 &      0 &&    125 &      0  & 33281\\
      0 &      0 &  33401 &&  33145 &      0 &    256 &&      0 &  33233  &   168\\ \hline
	\end{tabular}
\label{Tab4}
\end{table}

The main computational difficulty when maximizing the exact multinomial likelihood is that, for each local unit, the collection $\cg N_h$ has to be constructed and stored. To give an idea of how the size of $\cg N_h$ increases with the sample size, consider that in the $3\times 3$ case $\cg N_h$ contains about one hundred thousands tables with $n=100$, half a million for $n=150$ and over a million with $n=200$. Hence, at the moment, the method could not be applied to real life electoral data where the average size of polling stations is about 800. One possible solution under investigation is whether there exist meaningful subsets of $\cg N_h$ which, though of smaller size, lead to a close approximation of the score vector and the information matrix..

%
\section*{Appendix A: the log-likelihood}\label{secA1}
To derive the log likelihood of the column totals, we need to compute the product of the conditional distributions by row and sum across all possible configurations of the $n_{ij}$ compatible with the marginal totals. To condense notations, let $G(\bm  N)$ = $\sum_i\sum_j\log[\Gamma(n_{ij}+1)]$. The derivation below uses the following steps:
\begin{itemize}
\item start from the plain likelihood and than exponentiate its logarithm;
\item introduce the parametrization and deal with the numerator separately from the denominator;
\item note that the denominator and the terms in $\phi_j$ in the numerator depend only on marginal frequencies thus are constant across the elements of $\cg N$ and may be taken outside the summation;
\item when bringing out terms constant in $\cg N$, note that these terms are multiplicative factors when taken out of the exponential but become additive after applying the outer logarithmic operator.
\end{itemize}
\begin{align*}
L &= \log \sum_{\bm N\in\cgl N} \prod_{i=1}^R\left( n_{i0}!\prod_{j=1}^C \left[p_{ij}^{n_{ij}}/n_{ij}!\right] \right)
\propto \log \sum_{\bm N\in\cgl N} \exp \left[\sum_{i=1}^R\sum_{j=1}^C (n_{ij}\log(p_{ij})-
\log\Gamma(n_{ij}+1)) \right]\\
&=\log \sum_{\bm N\in\cgl N} \exp\left[\sum_{i=1}^R \sum_{j=1}^C n_{ij} (\phi_j+\lambda_{ij}) + G(\bm N) \right] -\sum_{i=1}^R n_{i0}\log\sum_{j=1}^C \exp(\phi_j+\lambda_{ij}) \\
&= \sum_{j=1}^{C-1} n_{0j} \phi_j - \sum_{i=1}^R n_{i0}\log\sum_{j=1}^{C-1} \exp(\phi_j+\lambda_{ij}) +
\log \sum_{\bm N\in\cgl N} \exp\left[\sum_{i=1}^R \sum_{j=1}^C n_{ij}\lambda_{ij} + G(\bm  N)\right].
\end{align*}
\subsection*{The likelihood equations}
Note that
$$
\frac{\partial}{\phi_j}\log\exp(\phi_j+\lambda_{ij}) = \frac{\partial}{\lambda_{ij}} \log\exp(\phi_j+\lambda_{ij}) = p_{ij}.
$$
The first set of $C-1$ equations are obtained by differentiating the first and second term of $L$ with respect to $\phi_j$:
$$
n_{0j} = \frac{\partial}{\partial\phi_j}\sum_{i=1}^R n_{i0}\log \exp(\phi_j+\lambda_{ij}) =
\sum_{i=1}^R n_{i0} p_{ij}, \quad j=2,\dots , C.
$$

The $\lambda_{ij}$ parameters appear in the second and third term of the log-likelihood, differentiating the second term gives
$$
\frac{\partial}{\partial\lambda_{ij}} \left(- \sum_i n_{i0}\log \sum_{j=1}^{C} \exp(\phi_j+\lambda_{ij})\right) = -n_{i0} p_{ij};
$$
by differentiating the last term by the chain rule, we obtain
\begin{align*}
& \frac{\partial}{\partial \lambda_{ij}}  \log \sum_{\bm N\in\cgl N}  \exp\left[\sum_{i=1}^R\sum_{j=1}^C n_{ij}\lambda_{ij} - G(\bm  N)\right] =\\
&\sum_{\bm N\in\cgl N} n_{ij} \frac{\exp\left[\sum_{i=1}^R\sum_{j=1}^C n_{ij}\lambda_{ij} + G(\bm  N)\right]}
 {\sum_{\bm N\in\cgl N} \exp\left[\sum_{i=1}^R\sum_{j=1}^C n_{ij}\lambda_{ij} + G(\bm  N)\right]} = M_{ij}
\end{align*}
where the $M_{ij}$ are the conditional expectation of the internal frequencies conditionally on the columns totals. The likelihood equations may be written as
$$
n_{0j} = \sum_{i=1}^R n_{i0} p_{ij},\quad  M_{ij} = n_{i0} p_{ij},\quad i=1,\dots,R,\: j=1,\dots,C.
$$

\bibliography{Plack}
\bibliographystyle{spbasic}
\end{document}